\documentstyle[psfig]{elsart}
\begin{document}
\begin{frontmatter}
\title{Ordering States with Entanglement Measures}
\author[London]{S. Virmani}
\author[London]{M. B. Plenio}
\address[London]{Optics Section, Blackett Laboratory, Imperial College, 
London SW7 2BZ, UK}
\begin{abstract}
We demonstrate that all good asymptotic entanglement measures are either 
identical or place a different ordering on the set of all quantum states.
\end{abstract}
\end{frontmatter}
\section{Introduction} 
The term entanglement has been used in many different ways since its
inception almost fifty years ago. Although it was originally used to 
describe the quantum correlations of the EPR paradox and Bell's inequalities,
it has recently come to be regarded as a resource that can be used to
perform various quantum information processing and communication tasks
\cite{Plenio1,preskill}. This more application-orientated aspect of 
entanglement has led to important questions regarding our ability to 
manipulate entanglement locally. Practically, we may never have sources
that generate perfect entanglement, and therefore it is necessary to explore
our ability to bring (using only local operations and classical
communication) diluted forms of entanglement into a more 
concentrated form, such as singlet states. This question has led to the development of entanglement 
purification methods such as those in \cite{Gisin,Bennett1,Bennett2}. Attempts 
to bound the efficiency of such procedures have then given rise to the concept 
of entanglement measures, which attempt to quantify the amount of entanglement 
in a given state. Indeed, a variety of entanglement measures have been proposed 
in recent years. Examples of various proposals can be found in 
\cite{Bennett3,Vedral1,Vedral2,Wootters1}. 

There are two main approaches towards
entanglement measures. A practical approach tackles such issues as to how much 
entanglement is required to generate a given state \cite{Wootters1} or how much 
entanglement can be distilled from it \cite{Bennett3}, whereas a more axiomatic 
approach first formulates conditions any reasonable entanglement measure has 
to satisfy \cite{Bennett3,Vedral1,Vedral2,Horodecki1}, and then constructs abstract 
measures which satisfy them \cite{Vedral1,Vedral2}. While there is agreement 
on a number of basic properties, such as the non-increase of entanglement under 
local operations and classical communication (LQCC), it would be interesting to 
see what other `natural' conditions one could add in order to narrow down the 
set of `acceptable' measures. In a recent paper, M.,P. \& R. Horodecki \cite{Horodecki1},
for example, considered a set of reasonable postulates in which a key ingredient 
was the assumption that the distillable entanglement is a well-defined measure of
entanglement. Starting out from these postulates, they deduced a number of useful
properties that any other entanglement measure has to satisfy.  

Here we would like to ask the following question. What are the consequences if 
we demand that any good entanglement measure should generate the same `ordering' on the 
set of density operators? To be precise, two entanglement measures $E_1$ and $E_2$ 
are said to generate the same order, if, for all density operators $\rho_1$
and $\rho_2$ we have that
\begin{equation}
	E_1(\rho_1) \le E_1(\rho_2) \Leftrightarrow E_2(\rho_1) \le E_2(\rho_2)	\;\; .
	\label{definition}
\end{equation}
In this paper we show that this requirement is indeed a very strong one. In fact, we 
show that for most of the entanglement measures which have been proposed, this requirement 
can only be satisfied if the two measures are identical. A consequence of this result 
is that if the distillable entanglement, the entanglement of formation, and the relative
entropy of entanglement are not identical, then they generate different orderings on 
the set of density operators.
This means that we cannot conclude that the distillable entanglement of $\rho_1$ is 
smaller than that of $\rho_2$ just because the entanglement of formation of $\rho_1$ 
may happen to be smaller than that of $\rho_2$. 

While questions like this have previously
been investigated numerically for special entanglement measures \cite{Eisert}, our
analytical reasoning applies to all measures that reduce to the entropy of entanglement 
for pure states. As it is commonly accepted that the unique measure of entanglement
for pure states is the entropy of entanglement \cite{Popescu,Vidal}, our argument 
applies to all good entanglement measures.

This paper is organized as follows: in the following section we formalize the conditions 
under which our reasoning applies, and briefly present some important measures which satisfy 
them. Then we present the proof of the theorem, and finally we discuss some of the implications 
of our result.

\section{Review of some Entanglement Measures}

In order for our argument to apply, we only require the property that our entanglement 
measures are equal for a set of states $C$, such that the set of entanglement values on 
$C$, $E(\rho \in C)$, is a dense subset of the set of all entanglement values. This 
characteristic is true for all the entanglement measures listed below (and many more), 
as they are all equivalent to a particular continuous function on the set of pure states. 
Given a pure state $|\psi \rangle$, they all define the entanglement as being 
$E(|\psi \rangle \langle \psi|) = S(tr_{A}(|\psi \rangle \langle \psi|))$, 
where $S(\rho)$ is the standard von Neumann entropy and $tr_{A}(...)$ denotes
the partial trace. Indeed,
as the entropy of entanglement is essentially the unique measure of entanglement for pure 
states \cite{Popescu,Vidal}, our reasoning applies to all entanglement measures. 
Some of the more notable candidates are briefly described as follows.

1. Entanglement of Formation, $E_{F}(\rho)$ \cite{Bennett3,Wootters1}. This measure is defined as:

\begin{equation}
	E_{F}(\rho) = min_{\rho =  
	\sum_{i} p_{i} |\psi_{i} \rangle \langle \psi_{i}|} \sum_{i} p_{i} E(|\psi_{i} \rangle \langle \psi_{i}|) \; .
	\label{form}
\end{equation}
Since this measure was first introduced, a slight modification was suggested in \cite{tot}. 
There it was proposed that the total entanglement of formation should be 
\begin{equation}
	E_{F}^{\mbox{tot}}(\rho) = lim_{n \rightarrow \infty} \frac {E_{F}(\rho^{\otimes n})} {n} \; ,
	\label{total}
\end{equation}
where $\rho^{\otimes n}$ is the $n$-fold tensor product of $\rho$.
The reasoning behind this is that it is not known whether $E_{F}(\rho)$ is additive under 
tensor products. It is therefore possible that the total entanglement of formation 
$E_{F}^{\mbox{tot}}(\rho)$ is lower than $E_{F}(\rho)$, 
and hence would be the appropriate measure of the number of singlets required to generate $\rho$ 
asymptotically. The physical interpretation of the total entanglement of formation is that 
if Alice and Bob start out with an asymptotically large supply of singlet states, this measure 
determines the minimum number of singlet states required to generate each copy of $\rho$ using LQCC.

2. Distillable Entanglement, $E_D(\rho)$ \cite{Bennett3,Wootters1}. If Alice and 
Bob start out with asymptotically many copies of $\rho$, this measure is defined as the maximum 
number of output singlet states they can obtain using LQCC per input copy of $\rho$.

3. Relative Entropy of Entanglement, $E_{R}(\rho)$ \cite{Vedral1,Vedral2,Rains1}. This measure 
is defined as 
\begin{equation}
	E_{R}(\rho) = min_{\sigma\in{\cal D}} S(\rho||\sigma) 
	= -S(\rho) + min_{\rho\in{\cal D}} \mbox{ tr} (-\rho \mbox{log} (\sigma)) \; ,
\end{equation}
where ${\cal D}$ is either the set of separable states \cite{Vedral1,Vedral2}, the set of PPT (positive partial transpose)
states \cite{Rains2}, or the set of non-distillable states \cite{Vedral3}. This measure cannot increase under LQCC and is known to be an upper bound 
on the distillable entanglement $E_D(\rho)$. It has proved very useful as in some cases it is an 
achievable upper bound to $E_D(\rho)$ \cite{Rains2}, a function which is difficult to calculate 
in general. In addition, $E_{R}(\rho)$ is known \cite{Vedral2,Rains2} to lie between the entanglement of formation as defined
in Eq. (\ref{form}), and the distillable entanglement $E_D(\rho)$.

\section{The Proposed Measures are either Equivalent or do not Possess the same Ordering}

We will prove our main result for just two measures of entanglement, $E_{F}(\rho)$ and $E_{D}(\rho)$, 
as the proof is identical for all others. As stated earlier, these measures reduce to $S(\rho_{A})$ 
for the pure states (in fact, all good asymptotic entanglement 
measures should reduce to this form on the pure states \cite{Popescu,Vidal}). As this is 
a continuous function of the eigenvalues of $\rho_{A}$, and completely covers the range of possible values 
of entanglement, it is possible to find a pure state corresponding to any value of entanglement for 
any of the measures. Now consider a general mixed state $\rho$, with entanglement of formation 
$E_{F}(\rho)$. Pick two pure states $|\phi\rangle$ and $|\psi\rangle$ with entanglements of formation 
$E_F(|\phi\rangle) = E_{F}(\rho)+\epsilon$ and $E_F(|\psi\rangle) = E_{F}(\rho)-\epsilon$ respectively
for some $\epsilon>0$. 
We will start out by assuming that both $E_{F}(\rho)$ and $E_D(\rho)$ place the same ordering on states. As 
$E_{F}(|\phi\rangle) \geq E_{F}(\rho) \geq E_{F}(|\psi\rangle)$, this would require that
\begin{equation}
	E_{D}(|\phi\rangle) \geq E_{D}(\rho) \geq E_{D}(|\psi\rangle)
\end{equation}
but as $E_{D}(|\phi\rangle) = E_{F}(|\phi\rangle)$ and $E_{D}(|\psi\rangle) = E_{F}(|\psi\rangle)$ 
due to equivalence on the pure states, we would hence require that:
\begin{equation}
	E_{F}(\rho) + \epsilon \geq E_{D}(\rho) \geq E_{F}(\rho) - \epsilon \;\; .
\end{equation}
Taking $\epsilon$ to zero we see that 
\begin{equation}
	E_{D}(\rho) = E_{F}(\rho)
\end{equation} 
for all states $\rho$, if we require both measures to place the same ordering on all states. 
This conclusion would be the same regardless of which measures we choose to compare, as long 
as they coincide on pure states.

\section{Discussion and Conclusions}

We have shown that we are faced with a simple choice: the equivalence and continuity of all 
measures on the pure states forces them to either be identical or induce a different ordering 
in general. This has interesting repercussions for investigations of entanglement. As most 
entanglement measures are very difficult to calculate in general, it might have been hoped 
that calculating a more tractable measure would allow us to at least work out the correct 
ordering for other measures. The reasoning presented here acts as a caveat against such an 
approach. 

There are also repercussions as to whether there is a unique asymptotic measure of entanglement 
for mixed states. It is already known \cite{pure} that there is no single number which 
characterises finite transformations of pure states as effectively as $S(\rho_{A})$ does in 
the infinite case. It has also recently been demonstrated \cite{Horodecki3} that the total entanglement of formation $E_{F}^{\mbox{tot}}$ given in Eq. (\ref{total}) is in some cases strictly greater than the distillable entanglement for mixed states.
Connecting this result with our work hence implies that $E_{F}^{\mbox{tot}}(\rho)$ does $not$ give the same ordering as $E_{D}(\rho)$, which in itself is a very surprising conclusion. 

\ack This work was supported by EPSRC, The Leverhulme Trust, the European Union and has benefitted
from a conversation with W.K. Wootters during the joint workshop of the ESF-QIT programme
and the Newton institute in Cambridge, July 1999. S.V.
would also like to thank Ben Tregenna for some very helpful discussions, and Daniel Jonathan for
demonstrating a great deal of patience over the past year. We would also like to thank Peter Knight for reading through the manuscript, and John Calsamiglia for pointing out some inaccuracies in it.

\end{document}